\documentclass[onecolumn,showpacs,preprintnumbers,superscriptaddress]{revtex4-2}
\usepackage{graphicx,color}
\usepackage{float}
\usepackage{amsmath}
\usepackage{bm}
\usepackage{braket}
\usepackage{booktabs}
\usepackage[pdftex,colorlinks=true, linkcolor = blue, citecolor=blue,urlcolor=blue, bookmarksnumbered=true, bookmarksopen=true]{hyperref}
\usepackage[normalem]{ulem}
\begin{document}
\newcommand{\joec}[1]{{\color{red} #1}}
\newcommand{\cyantext}[1]{{\color{cyan}#1}}

\title{State preparation and symmetries}

\author{Ivana Mih\'alikov\'a}
\affiliation{Institute of Physics of Materials, Czech Academy of Sciences, Brno 616 00, Czech Republic}
\affiliation{Institute of Physics, Slovak Academy of Sciences, Bratislava 84511, Slovakia; Matej Bel University, N\'arodná ulica 12, Bansk\'a Bystrica, 97401, Slovakia}
\affiliation{Department of Condensed Matter Physics, Faculty of Science, Masaryk University, Brno 611 37, Czech Republic}
 \affiliation{Los Alamos National Laboratory, Theoretical Division, Los Alamos, New Mexico 87545, USA}

\author{Joseph Carlson}
 \affiliation{Los Alamos National Laboratory, Theoretical Division, Los Alamos, New Mexico 87545, USA}

 \author{Duff Neill}
  \affiliation{Los Alamos National Laboratory, Theoretical Division, Los Alamos, New Mexico 87545, USA}

 \author{Ionel Stetcu}
 \affiliation{Los Alamos National Laboratory, Theoretical Division, Los Alamos, New Mexico 87545, USA}

\begin{abstract}

We demonstrate the importance of symmetries in Variational Quantum Eigensolver (VQE) algorithms to prepare the ground or specific low-lying states of quantum Hamiltonians. We examine  two spin problems, one with random all-to-all couplings inspired by neutrino flavor evolution in supernovae, and the standard Heisenberg spin Hamiltonian on a $4 \times 3$ lattice.
The~neutrino Hamiltonian has the total spin $J$ and third component $J_{\rm{z}}$ as its only symmetries.  The~Heisenberg model has these symmetries plus translational invariance and reflection symmetry.

We demonstrate that 
the convergence of variational methods is dramatically improved by keeping all symmetries. In both cases a nearly exact solution can be obtained in cases where standard unconstrained variational algorithms fail. Since variational algorithms can use standard Trotter steps as part of the optimization, allowing additional correlations that obey all the symmetries of the Hamiltonian will speed convergence of variational
algorithms. This will lead to faster convergence than standard projection algorithms. 
\end{abstract}
\pacs{}
\preprint{LA-UR-25-29124}
\maketitle

\section{Introduction}

Quantum ground-state preparation requires good classical product-state ansatz~\cite{PRXQuantum.5.040339, Bauer2020ChemRev,PRXQuantum.6.020327}. Ideally, the ansatz should have large overlap with the true ground state of the system and should be simple enough to be easily prepared on a quantum computer, if possible only involving single-qubit gates. 
However, for a given target Hamiltonian, it may be the case that only poor guarantees can be made for the initial product state overlap, and even finding the optimal classical product state can be shown to be a hard problem, if the system is a classical spin-glass, for instance~\cite{2016CMaPh.342...47B,2024arXiv241008322K,https://doi.org/10.4230/lipics.itcs.2025.63,anschuetz2024boundsgroundstateenergy, Barahona_1982}. The guarantees are generally a function of the locality and dimensionality of the Hamiltonian, where we use locality in the computer science sense of number of sites that can interact via a single term in the Hamiltonian, and dimensionality to count the number of sites that a given site can interact with. For problems with a spatial lattice, this counting dimensionality is directly correlated with the spatial dimensionality of the lattice. Generally, as the dimensionality increases, one can expect product states to produce an ever better overlap with the ground state. The underlying intuition is the monogamy of entanglement: each site is pushed to being entangled with the sites that it can interact with, but must share that entanglement in a roughly equal fashion according to the dimensionality of the Hamiltonian. 

For large lattices in low dimensions, the entanglement can be high. Variational quantum algorithms can be useful, though they often suffer from the 'barren plateau' issue~\cite{bp1,bp2,bp3,bp4,bp5,PRXQuantum.2.040316}. In many directions the optimized variational function has very small gradients
with respect to changes in the (many) variational parameters~\cite{Cerezo:2021,federov:2022,vqe,vqe1, PhysRevC.105.064308}. However, because barren plateaus describe only the average landscape, regions with non-vanishing gradients can still be found when the circuit is initialized sufficiently close to a good minimum~\cite{grimsley2023adaptive,PhysRevA.106.L060401,Park2024hamiltonian,PhysRevApplied.22.054005, PhysRevResearch.2.043142}. Reducing the number of variational parameters can produce useful results, though sometimes at the cost of reduced accuracy.

An important consideration for many problems of physical interest is the presence of either exact or approximate symmetries in the problem~\cite{Gibbs2025, 10.1063/1.4998921, Yen2019ExactAA, PRXQuantum.2.010323, Lacroix2023}. In the limit of exact symmetries  the Hilbert space will break up into non-interacting subspaces, each subspace being specified by a set of quantum numbers that characterize an irreducible representation of the symmetry group. Constructing states that belong to a specific irreducible representation in general requires entanglement. Take the case of SU($2$) as a concrete example of a symmetry group, and a Hilbert Space composed from $N$ qubits. Then it is simple to see that to build a state with total spin of zero, one can pair each spin, and put the pairs into a locally spin zero state, anti-symmetric under interchange. Such symmetry-adapted product states can serve as warm starts for quantum algorithms, where the initial state is prepared with respect to the symmetry of the problem~\cite{truger2024warmstartingvqeapproximatecomplex,Egger2021warmstartingquantum,chai2024structureinspiredansatzwarmstart,PRXQuantum.6.010317,Sheikh_2021}.


In this work, we examine, for small systems, how effectively a product-state ansatz can approximate the true ground state of symmetric Hamiltonians, and how variational quantum algorithms can enhance this process. We then explore the extent to which projection algorithms alone can improve the overlap of such product states with the total spin-zero subspace. Following this projection, we apply a restricted variational quantum algorithm specifically designed to preserve the underlying symmetries of the Hamiltonian. We focus on the case for an underlying SU($2$) symmetry, comparing an all-to-all coupled system inspired from neutrino physics~\cite{PRXQuantum.4.027001, PRXQuantum.5.037001} (where the symmetry derives from the approximate flavor symmetry of the neutrino sector of the Standard Model under neutral current interactions) to the standard magnetic Heisenberg model in 2 dimensions. For the neutrino-inspired Hamiltonian, it is simple to write down the specific product state that is supposed to dominate the ground state in the classical limit \cite{neill2024scatteringneutrinosspinmodels} as the number of neutrinos becomes large, a state called the flavor-momentum locked state. For the Heisenberg model, we use the anti-ferromagnetic N\'eel state as the initial product state~\cite{RevModPhys.63.1}.

\section{Models and Algorithms}

\subsection{Spin models}

The Heisenberg spin model \cite{Heisenberg1928} is a fundamental framework in statistical mechanics and condensed matter physics that describes the interactions between spins arranged on a lattice. In this work, we consider two different Hamiltonians. One is the
standard spin $1/2$ Heisenberg Hamiltonian on a $2$D square lattice. This Hamiltonian is given by
\begin{equation}
H =   \sum_{<{\rm{ij}}> } \sigma_{\rm{i}} \cdot \sigma_{\rm{j}},
\end{equation}
where the sum runs over nearest-neighbor pairs on the lattice.
For illustrative purposes we use a $2$D square lattice with periodic boundary conditions, but similar techniques could be used on other lattices including
those that cannot be simulated exactly by classical computers with Quantum Monte Carlo methods.
This Hamiltonian has several symmetries.  The total spin and its projection on an axis are conserved, and there are translational symmetry and mirror symmetries in the $x$- and $y$-planes.

We also consider the Heisenberg-like spin Hamiltonian  including all-to-all random interactions between pairs of
spins.  This model has often been used to study neutrino dynamics in supernovae~\cite{neill2024scatteringneutrinosspinmodels}. In this case, the Hamiltonian is

\begin{equation}
    H = \sum_{{\rm{i}}<{\rm{j}}} (1 - \cos{\theta_{\rm{ij}}}) \  \sigma_{\rm{i}}\cdot \sigma_{\rm{j}}.
\end{equation}
The term $ (1 - \cos{\theta_{\rm{ij}}})$ represents a coupling strength or interaction coefficient between the ${\rm{i}}$-th and ${\rm{j}}$-th SU($2$) spin in the system, where $\theta_{\rm{ij}}$ is an angle parameter that influences the interaction, and ${\sigma}_{\rm{i}}$ and ${\sigma}_{\rm{j}}$ are Pauli matrices representing the spin operators at sites ${\rm{i}}$ and ${\rm{j}}$, respectively. In our model, $\cos{\theta_{\rm{ij}}}$ is randomly chosen from the range $[-1, 1]$, ensuring that the coupling strength is always non-negative and varies smoothly from $0$ to $2$. 
%

This Hamiltonian again conserves the total spin $J$ and its third component $J_{\rm{z}}$. However, there are no apparent additional symmetries.

\subsection{Initial states}

In our variational quantum simulations of these spin models, we start with the tensor product of individual spin coherent states for $N$ spins~\cite{neill2024scatteringneutrinosspinmodels} as an initial state.
These states are very simple product states that are easy to prepare on a quantum computer.
For the standard Heisenberg model we use the symmetrized N\'eel state
\begin{equation}
    \ket{\psi_{\rm{i}}} = \frac{1}{\sqrt{2}} [ \prod_{\rm{i}} \sigma_{\rm{x}} (i)^{x+y} + \prod_{\rm{i}} \sigma_{\rm{x}} (i)^{x+y+1}]  \prod_{\rm{i}} |0 \rangle_{\rm{i}}. 
\end{equation}
This state is unique for a square lattice with even lengths in both directions. All nearest neighbors have
expectation value -1 for that term in the Hamiltonian. For the $4 \times 3$ lattice
considered here, we also use this state, although the periodic boundary conditions imply that the pairs
wrapping around the lattice in one direction have a positive expectation value. 

For the all-to-all coupled neutrino model, we employ a product state. In this scenario, we use the state outlined in Ref.~\cite{neill2024scatteringneutrinosspinmodels}, where the spin of each particle is aligned to the momentum.
The state is
\begin{equation}
\psi_{\rm{i}} = \prod_i R(i) | 0 \rangle,
\end{equation}
with the rotation $R$ defined so that its expectation values are parallel to the momenta used to 
define the Hamiltonian. 

In other words, the state can be written as

\begin{equation}
\label{duffs_state}
    \ket{\psi_{\rm{i}}} = \mathop{\bigotimes}\limits^{N}_{{\rm{i}}=1} \exp\left(i \pi {k}_{\rm{i}} \cdot \frac{{\sigma}_{\rm{i}}}{2}\right) \ket{0}_{\rm{i}},
\end{equation}

\noindent where ${k}_{\rm{i}}$ is parallel to ${\sigma}_{\rm{i}}$, and ${k}_{\rm{j}}$ is parallel to ${\sigma}_{\rm{j}}$.

The vectors ${k}$ are defined as

\begin{equation}
      {k}_{\rm{i}} = \cos(\phi_{\rm{i}}) \sin\left(\frac{\theta_{\rm{i}}}{2}\right) {e}_{\rm{x}} + \sin(\phi_{\rm{i}}) \sin\left(\frac{\theta_{\rm{i}}}{2}\right) {e}_{\rm{y}} + \cos\left(\frac{\theta_{\rm{i}}}{2}\right) {e}_{\rm{z}},
\end{equation}

\noindent the azimuthal angle $\phi$ is randomly chosen from the range $[0, 2\pi]$. These vectors represent spin directions in spherical coordinates, capturing the randomized nature of the system's initial state.

This initial state represents a product of single-spin states, where each spin is rotated from the reference state $\ket{0}_{\rm{i}}$ by an operator $\exp(i \pi {k}_{\rm{i}} \cdot {\sigma}_{\rm{i}} / 2)$, where ${k}_{\rm{i}}$ is a unit vector associated with each neutrino, representing its momentum, and ${\sigma}_{\rm{i}}$ is the Pauli spin vector operator acting on qubit ${\rm{i}}$. This tensor product state can be considered as a mean-field-like state where each spin aligns with its corresponding momentum vector. It approximates the ground state of the system in the classical limit and provides a semiclassical interpretation of the quantum dynamics. 

To implement this state on quantum computer, we used the fact that any single-qubit operator can be decomposed into a sequence of three rotations about the $z$- and $y$-axes. Specifically, we used $R_{\rm{z}}R_{\rm{y}}R_{\rm{z}}$ decomposition that provides a symmetric and standard way to decompose arbitrary rotations, reflecting the Euler angle decomposition

\begin{equation}
\begin{aligned}
\label{psi}
\ket{\psi_{\rm{i}}} = \exp{( -i \alpha_{\rm{i}} Z)} \exp{(-i \beta_{\rm{i}} Y)} \exp{( -i \gamma_{\rm{i}} Z)} \ket{0}_{\rm{i}} = R_{\rm{z}}(\alpha_{\rm{i}}) R_{\rm{y}}(\beta_{\rm{i}}) R_{\rm{z}}(\gamma_{\rm{i}}) \ket{0}_{\rm{i}}.
\end{aligned}
\end{equation}

\subsection{Spin Projection Algorithm}

One of the applications of the projection algorithm introduced in Ref.~\cite{PhysRevC.108.L031306} was to prepare states with good symmetries. However, for projection on total spin $J=0$ it is more efficient to use a simplified algorithm \cite{PhysRevC.110.064003} that requires more measurements but avoids deep quantum circuits with complex Pauli strings and Trotterization. For the Heisenberg model this approach simplifies further and consists of repeated projections on total spin projection zero using the operator $J_{\rm{z}}$ (which involve a measurement), followed by a single rotation by $\pi/2$ around the $x$ axis.  This significantly reduces circuit depth and makes the approach more practical for today's quantum hardware. To perform the projection on $J = 0$, the algorithm starts with an initial quantum state, which is generally a superposition of states with different values of total angular momentum $J$ and their corresponding magnetic projections $M$. The goal is to remove all components that do not belong to the $J = 0$ subspace.

One of the key steps in this process is the application of unitary evolution followed by a measurement of an ancilla to eliminate unwanted components. The state evolution under $J_{\rm{z}}$ is given by

\begin{equation}
\begin{aligned}
\label{Jz}
\ket{\psi(t_{\rm{i}})} &= \exp (-i t_{\rm{i}} J_{\rm{z}} \otimes Y_{\rm{a}}) \ket{\psi} \otimes \ket{0}_{\rm{a}} \\
&= \cos(J_{\rm{z} }t_{\rm{i}}) \ket{\psi} \otimes \ket{0}_{\rm{a}} + \sin(J_{\rm{z}} t_{\rm{i}}) \ket{\psi} \otimes \ket{1}_{\rm{a}}.
\end{aligned}
\end{equation}

Choosing a proper series of times $t_{\rm{i}}=\pi/2^i$ ($i>0$), repeated measurements of the ancilla in state $|0\rangle_a$ projects on many-body states with $J_{\rm{z}}=0$. However, because states with $J>0$ also have a $J_z=0$ component, the algorithm must continue to filter out those components. A rotation around the $x$ (or $y$) axis with an angle $\varphi \neq 0$ or $\pi$ reshuffles contributions from $J>0$, $J_{\rm{z}}=0$ states to states with $J>0$, $|J_{\rm{z}}|>0$, and hence an additional projection as above reduces the weight of the $J>0$ states, which become zero after a relatively small number of iterations that involve projection followed by rotation around the $x$ or $y$ axis by $\pi/2$ (the last-step rotation is not necessary) \cite{PhysRevC.110.064003}. The exact number of iterations depends on the contamination with states with $J>0$, but in practice we needed to perform about $11$ iterations, somewhat better than in Ref. \cite{PhysRevC.110.064003} where the number of iterations for nuclear physics problems was around $40-50$.  
While in other cases, like in nuclear physics for even-even nuclei, this algorithm could be used to project on $J=0$ states, increasing the gap and hence allowing for a more efficient state preparation via projections \cite{PhysRevC.108.L031306}, in the current investigation we use it to efficiently project onto states with
total $J=0$ and $J_{\rm{z}} = 0$ (equal numbers of spin up and down), and then continue with a variational approach.

\subsection{Variational Ansatz}


In our VQE calculations, we utilize an ansatz that incorporates the swap operator $\hat{\rho}_{\rm{ij}}$, which exchanges the quantum states of two qubits. This operator has the advantage of conserving both the total and $\rm{z}$-component of the spin of the pair, and hence the whole system. The exponentiation of this operator, $\exp(i\theta \hat{\rho}_{\rm{ij}})$, generates entangling interactions that can be decomposed into a sequence of two-qubit Pauli interactions of the form $\exp(i\theta \sigma_{\rm{i}}^a \sigma_{\rm{j}}^a)$ for $a = X, Y, Z$, which are implemented using controlled-NOT (CNOT) gates and single-qubit rotations. The exponentiated swap operator is given by

\begin{equation}
\exp(i\theta \hat{\rho}_{\rm{ij}}) =
\begin{bmatrix}
e^{i\theta} & 0 & 0 & 0 \\
0 & \cos(\theta) & i \sin(\theta) & 0 \\
0 & i \sin(\theta) & \cos(\theta) & 0 \\
0 & 0 & 0 & e^{i\theta}
\end{bmatrix},
\end{equation}

\noindent where $\theta$ represents the variational parameter. The individual exponentiated Pauli interactions have the following forms

\begin{equation}
\exp(i\theta \sigma^X_{\rm{i}} \sigma^X_{\rm{j}}) = 
H_{\rm{i}} H_{\rm{j}} \text{CNOT}_{{\rm{i}} \to {\rm{j}}} R_{{\rm{i,z}}}(-\theta) \text{CNOT}_{{\rm{i}} \to {\rm{j}}} H_{\rm{i}} H_{\rm{j}}.
\end{equation}

\begin{equation}
\exp(i\theta \sigma^Y_{\rm{i}} \sigma^Y_{\rm{j}}) =  
H_{\rm{i}}^{Y} H_{\rm{j}}^{Y} \text{CNOT}_{{\rm{i}} \to {\rm{j}}} R_{\rm{i,z}}(-\theta) 
\text{CNOT}_{{\rm{i}} \to {\rm{j}}} H_{\rm{i}}^{Y\dagger} H_{\rm{j}}^{Y\dagger},
\end{equation}

\noindent where

\begin{equation}
    H_{\rm{i}}^{Y} = S_{\rm{i}} H_{\rm{i}} \sigma^Z_{\rm{i}} S_{\rm{i}},
\end{equation}

\noindent and

\begin{equation}
    H_{\rm{i}}^{Y\dagger} = \sigma^Z_{\rm{i}} S_{\rm{i}} \sigma^Z_{\rm{i}} H_{\rm{i}} \sigma^Z_{\rm{i}} S_{\rm{i}}.
\end{equation}

\begin{equation}
\exp(i\theta \sigma^Z_{\rm{i}} \sigma^Z_{\rm{j}}) = \text{CNOT}_{{\rm{i}} \to {\rm{j}}} R_{\rm{i,z}}(-\theta) \text{CNOT}_{{\rm{i}} \to {\rm{j}}}.
\end{equation}

By combining these exponentials, we obtain the full exponentiated swap operator decomposition

\begin{equation}
\label{operator}
\exp\left(i \theta \hat{\rho}_{\rm{ij}} \right) =  
 e^{i\theta/2} 
\exp\left(i \frac{\theta}{2} \sigma^X_{\rm{i}} \sigma^X_{\rm{j}} \right)
\exp\left(i \frac{\theta}{2} \sigma^Y_{\rm{i}} \sigma^Y_{\rm{j}} \right)
\exp\left(i \frac{\theta}{2} \sigma^Z_{\rm{i}} \sigma^Z_{\rm{j}} \right).
\end{equation}

Here, $e^{i\theta/2}$ represents an overall phase factor that does not affect the computational basis measurement results. This ansatz preserves the interaction structure while maintaining low circuit depth.

Arbitrary pair spin swaps are appropriate for the neutrino Hamiltonian as the spin symmetries are the only relevant ones. However, for the Heisenberg Hamiltonian, translational and reflection symmetry are also important.

We can enforce these symmetries approximately by first considering all nearest-neighbor pairs and assigning the same swap amplitude to each. Correlations can then be added for next-nearest-neighbor pairs, such as those separated by a distance of 2 in one direction, and so on. In each case, all pairs are equally spaced, which helps maintain approximate translational symmetry. However, the symmetry is not exact because not all pair operations commute, leading to small admixtures of states that break translational invariance.



\section{Numerical results}

We show results for the neutrino model with $12$ randomly coupled spins as described above and
for the $4 \times 3$ Heisenberg model that has many
additional symmeties. Our initial product states have energy distance to the ground state well within the bounds derived in Ref. \cite{2016CMaPh.342...47B}, which are 166.7 energy units to the ground state for the neutrino model and 42.4 for the Heisenberg model. We note that the normalization differences between our hamiltonians and that of Ref. \cite{2016CMaPh.342...47B}, which always normalizes by the total number of spins and the degree of the interaction graph of the hamiltonian, that is, the number of sites each site can be connected to.. This indicates our initial product-states are performing much better than the worst case guarantees for on how well a product state can approximate the ground state. Specifically, these bounds quantify how well a product-state can approximate the ground of a quantum hamiltonian, given no other considerations about the underlying physics. Roughly speaking, they bound the energy per site or qubit involved in the interaction, and per number of sites/qubits that a single site/qubit can interact with. For the neutrino interactions, this becomes a more strigent bound, as the number of sites that interact with a given site scales with the system size, while for the Heisenberg model, the bound becomes a constant as the system grows in size. 

The first step after creating the product state is to project onto the $J = 0$ subspace.  With a finite number of steps, the
convergence is not exact, but it is very rapid. 

\begin{figure*}[h]
    \centering
    \includegraphics[width=0.6\linewidth]{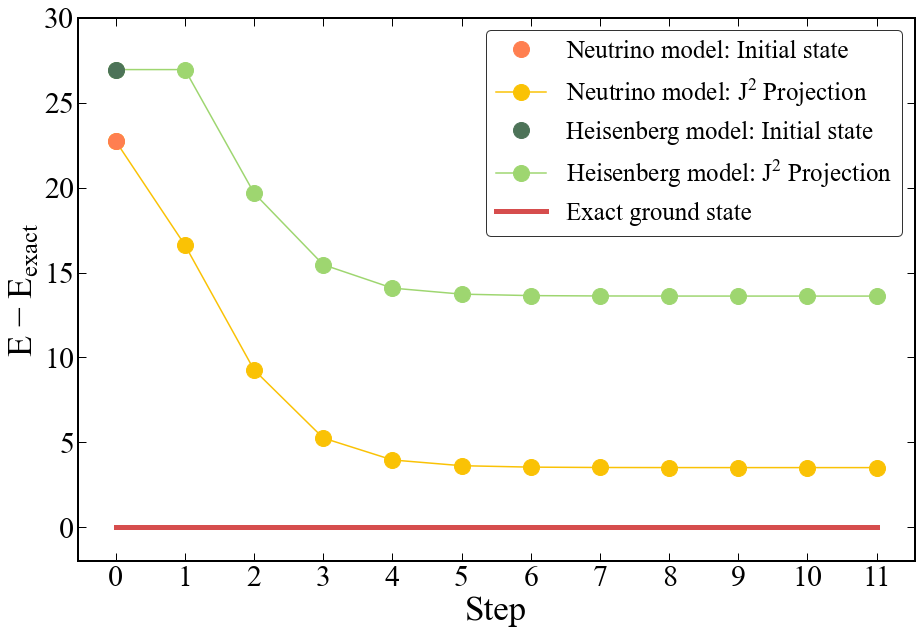}
    \caption{Energy difference ($E - E_{\rm{exact}}$) of the projected state at each step of the projection algorithm applied to the initial state of the Neutrino model, and the Heisenberg model. The unprojected initial state is shown as the first point at step $0$, while the subsequent points connected by solid lines represent the energy after each of the $11$ projection steps. The red line indicates the exact ground state energy.}
    \label{fig:projection}
\end{figure*}

The energy versus iteration step of the projection to spin zero is shown in Fig. \ref{fig:projection}.
For the neutrino case, the product state should become nearly exact for large $N$, projection to spin zero improves the initial state dramatically for $12$ spins, improving the energy approximately $80\%$ of the way to the exact.  For the Heisenberg case, there are additional symmetries important to get the ground state correct.  Projecting on good total spin (and $J_{\rm{z}}$) is important, but the spin projection 
alone improves the energy around $50\%$ to the 
ground state.

\begin{figure*}[h]
    \centering
    \includegraphics[width=0.9\linewidth]{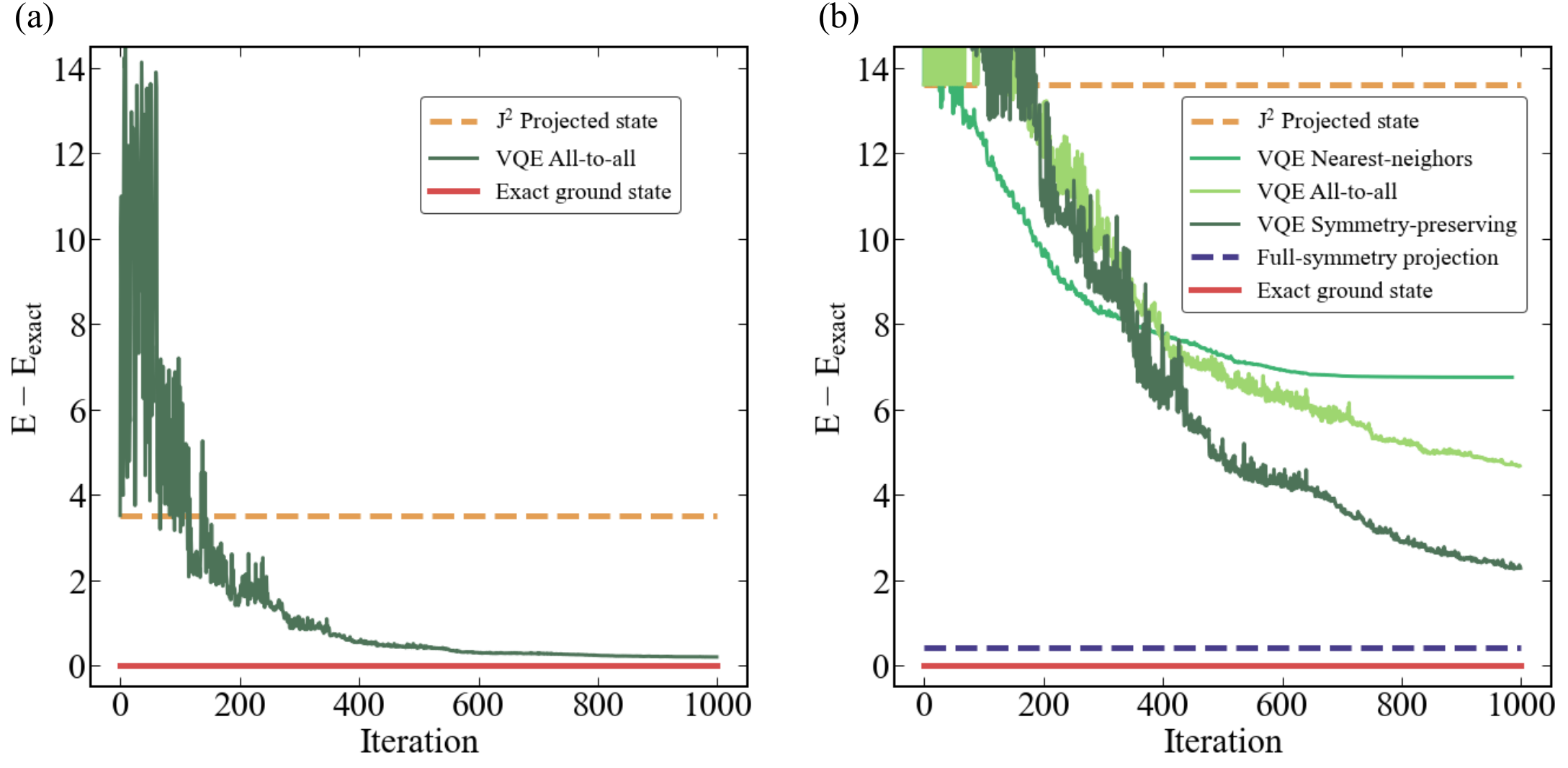}
    \caption{Swap convergence plots using the VQE algorithm for the Neutrino model (a), and the Heisenberg model (b) starting from the projected state obtained after $11$ steps of the projection algorithm represented by the orange dashed line. The red line is the exact ground state energy. The plots show the energy difference $E - E_{\rm{exact}}$. In both cases, the dark green curve shows the VQE energy as a function of iteration that preserve the symmetry of the model. In the Heisenberg case (b) additional strategies are shown: the green curve uses nearest-neighbor connectivity, and the olive-green curve uses all-to-all connectivity. Final translational/reflection symmetry projected state from VQE is shown as a blue dashed line.}
    \label{fig:convergence}
\end{figure*}

The next step in the algorithm is the VQE convergence, with results shown in Fig. \ref{fig:convergence}.  For the neutrino case, see Fig. \ref{fig:convergence}(a), we use random pair swaps as projection operators to construct the all-to-all connected ansatz for VQE optimization, projecting state into more entangled subspace. The swaps preserve the total $J=0$ character of the state.  This system has a fairly large gap to higher $J=0$ states, which also
improves the convergence. The final energy of the VQE projected state is $ E = - 47.46$ compared to the exact $-47.66$ (see Table \ref{table:energiesoverlaps}). 


\begin{table}[h]
\centering
\setlength{\tabcolsep}{12pt} 
\caption{Energy values during optimization for the Neutrino and Heisenberg models.}
\begin{tabular}{@{}ccccc@{}}
\toprule
\textbf{Optimization} & \textbf{Neutrino} & \textbf{Fidelity} & \textbf{Heisenberg} & \textbf{Fidelity} \\
\textbf{step}         & \textbf{energy} & & \textbf{energy} &  \\
\midrule
Initial product state              & $-24.93$ & $0.024$ & $-32.00$ & $0.085$ \\
$J^2$ Projection $11$ iterations   & $-44.15$ & $0.567$ & $-45.33$ & $0.296$ \\
VQE convergence                   & $-47.46$ & $0.988$ & $-56.66$ & $0.897$ \\
Translation/Mirror Projection     & --         &   --    & $-58.52$ & $0.985$  \\
Reference                         & $-47.66$ & $1.000$ & $-58.95$ & $1.000$ \\
\bottomrule
\end{tabular}
\label{table:energiesoverlaps}
\end{table}

For Heisenberg, in Fig. \ref{fig:convergence}(b), there are multiple symmetries. We compare results for three approaches: (i) a VQE with randomly ordered nearest-neighbor swaps, (ii) an all-to-all VQE, and (iii) a VQE that approximately conserves translation and mirror symmetry. For the last case, the algorithm uses the same parameters for all swaps between pairs of spins that are the same distance apart. One layer connects the nearest neighbors (vertical and horizontal), another connects the diagonal ones, and then the second nearest neighbors. We applied periodic boundary conditions to ensure that the pattern is consistent across the lattice. This setup respects the symmetries of the system and still preserve the total spin. 
We initialized the variational parameters (Eq.~\ref{operator}) in the ansatz according to the interaction type. For the first layer, which applies gates between nearest neighbors (vertical and horizontal), we set the parameter values to $0.01$. For the second layer, which connects diagonal neighbors, we used slightly larger values of $0.15$, and for the third layer, which connects second nearest neighbors, we again set the parameters to $0.01$. Small initial parameters for the nearest and second nearest neighbor layers $(0.01)$ keep the circuit close to the identity at the start of the optimization. The larger initial value for the diagonal layer $(0.15)$ gives more weight to diagonal interactions, which can introduce relevant correlations in the system.

(i) The nearest-neighbor VQE, retains the $J=0$ character of the ground state, but is difficult for the algorithm to find a translationally invariant state since it is using only local operators.  Thus, the final energy does not work as well as the
symmetry-preserving algorithms.  

(ii) The VQE all-to-all, again preserves the total spin, in principle it can also obtain a translationally and mirror invariant state but in practice it is difficult to find that.  It improves the longer distance correlations (entanglement) compared to the nearest-neighbor.

(iii) The best algorithm is the one that (approximately) preserves all symmetries. The symmetries are not exact because of non-vanishing commutators between different pairs at the same separation. This VQE yields an energy of $-56.66$ compared to the initial product state energy of $-32.00$ and to the spin projected N\'eel state energy of $-45.33$, while the exact value of energy is $-58.95$.

One can project on translational symmetry and mirror symmetries
in a manner very similar to the spin projection above.  In effect, we add a term to the Hamiltonian

\begin{equation}
\delta H = \sum_{{\rm{x, y}}} c_{\rm{x, y}} \prod_{(i,j)} T(j \rightarrow i).
\end{equation}
This term simultaneously translates each spin to a nearest-neighbor site in the $x$- or $y$-directions. The $x$- and $y$- terms
can have different coefficients. This essentially adds a harmonic oscillator term with $x$- and $y$-masses in the center of mass.
The spectrum is linear in energy (a~harmonic oscillator) and all
excitations can be projected out very quickly. The mirror symmetries can be treated similarly.

Projecting out exact translational and mirror symmetries from the
VQE state (the $k^2$ projection, where $k=(k_{\rm{x}},k_{\rm{y}})$ denotes the lattice momentum quantum numbers associated with translations in the $x$- and $y$-directions and the state is restricted to the translationally invariant $k=0$ sector) yields an energy of $-58.52$ compared to the exact $-58.95$, and is indicated by the horizontal dark blue dash-dotted line in Fig. \ref{fig:convergence}.  
The algorithm for such a projection is very similar to the spin projection described above.

\begin{figure*}[h]
    \centering
    \includegraphics[width=0.95\linewidth,clip]{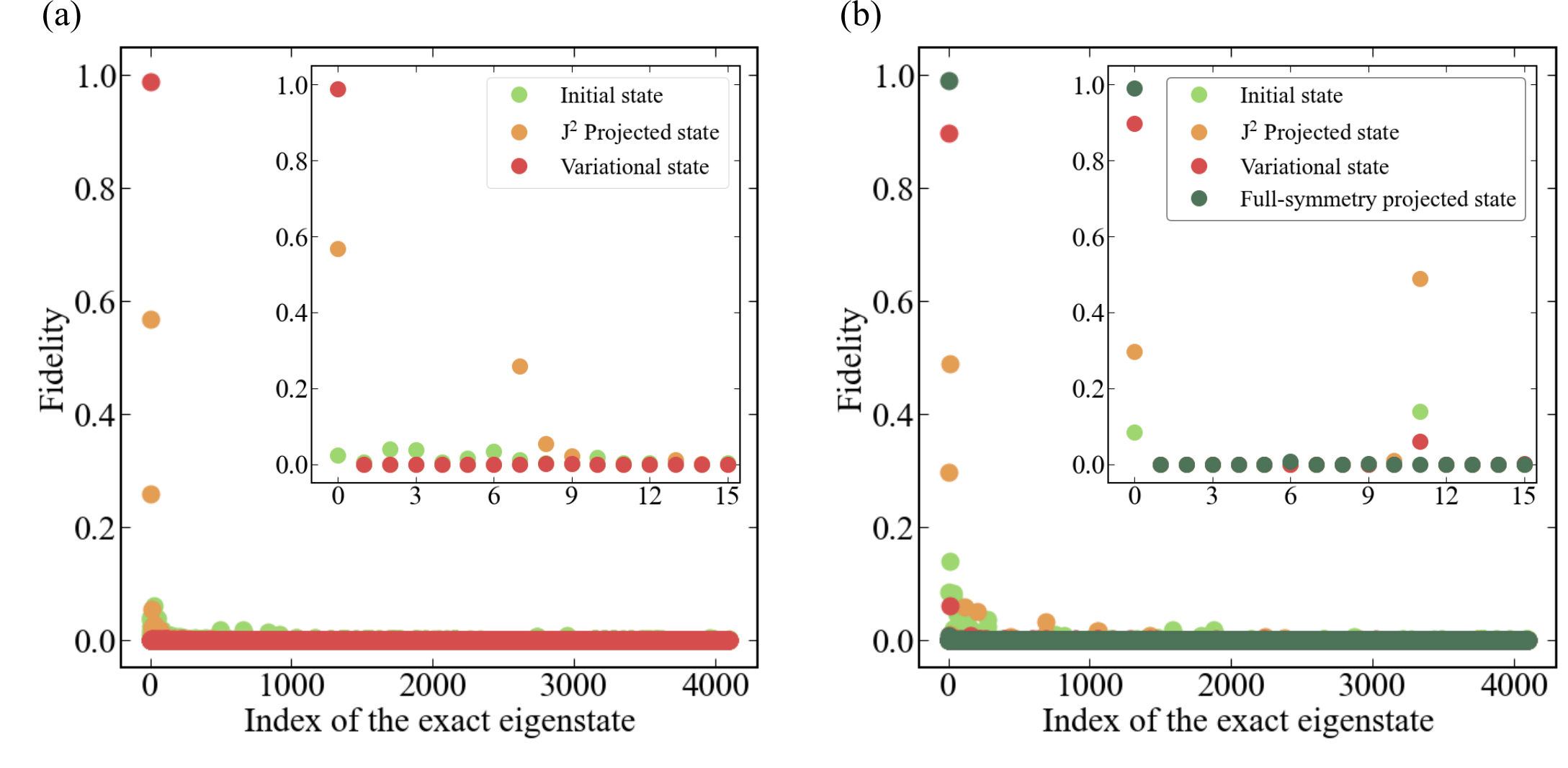}
    \caption{Fidelity of the prepared state with respect to the $4096$ classically computed eigenstates during optimization for the Neutrino model (a), and the Heisenberg model (b). The fidelity is computed as $|\braket{\psi | \phi}|^2$, where $\ket{\psi}$ is the prepared state and $\ket{\phi}$ is the reference state.}
    \label{fig:fidelities}
\end{figure*}

Figure \ref{fig:fidelities} shows the squared overlaps (fidelities) obtained by comparing each of our prepared states with all exact eigenstates of the corresponding Hamiltonian, which were obtained classically. The results are displayed for three prepared states: the initial state, the state obtained after projection onto the appropriate symmetry subspace, and the state obtained after VQE optimization. In the case of the Heisenberg model, the fourth state corresponds to the VQE optimized state after the additional symmetry projection.

In case of the neutrino model, Fig.~\ref{fig:fidelities}(a) shows the fidelity between the initial coherent product state and the exact eigenstates of the fully connected neutrino Hamiltonian. The overlap with the exact ground state is relatively low $2.4 \%$, and to improve it, we project the state onto the $J = 0$ subspace. The projection leads to a state whose maximum overlap matches the exact ground state $56.7 \%$. The plot further shows only minor contributions from the exact higher-energy states, confirming that the projection procedure effectively filters out components of the studied product state outside the target symmetry subspace. The fidelities of the final state obtained after the full variational optimization with the exact eigenstates of the neutrino Hamiltonian exhibit almost perfect overlap with the true ground state $98.8 \%$, while contributions from all higher-energy states are negligibly small. 

For the Heisenberg model, the overlap squared of the superposition of two N\'eel states with the true ground state, that is in the total spin $J=0$ subspace, is $8.5 \%$, see Fig.~\ref{fig:fidelities}(b). However, significant overlap is observed with two exact higher-energy states, which are $0.3 \%$ and $14.0 \%$. With the projection onto the total spin $ J = 0 $, we reduce the size of the state space and provide a better starting point for the VQE convergence. The overlap squared of the projected of the studied product state with the exact ground state is $29.6 \%$. Even after applying the projection algorithm for $11$ iterations, the resulting state still shows a nonzero overlap with exact higher-energy eigenstates. Despite this, the projection improves the overlap with the exact ground state and helps concentrate the state within the correct symmetry sector. The state obtained after VQE optimization has a very high overlap with the exact ground state, which is $89.7 \%$. Furthermore, this state can be exactly projected onto a state of good translational symmetry and good mirror symmetry in the $x$- and $y$- directions, yielding an overlap with the true ground state of greater than $98\%$, similar to the neutrino Hamiltonian.

In each case, the final VQE state has excellent overlap with the ground state, and projecting on the symmetries dramatically increases the gap. Further improvements by projection algorithms (\cite{PhysRevC.108.L031306} would be very effective.

It is not entirely surprising that these symmetries are so important.
The full Hilbert space for a $4 \times 3$ lattice is 2$^{12}= 4096$. Of these, $924$ have $J_{\rm{z}} = 0$,
restricting to $J=0$ as well yields $132$ states. Further restricting to translationally invariant states and mirror symmetry drastically reduces the number of relevant states to $9$. 

Similarly, the gap to the first excited
state increases dramatically when all symmetries are retained. The exact ground state is 
at $E=-58.95$, the first excited state with $J_{\rm{z}} = 0$ is at $-51.81$, the first with $J=0$ and $J_{\rm{z}} = 0$ is at $-46.72$.  The first excited state with all the same symmetries is at an energy of $-30.86$. The increased gap 
is extremely valuable in time projection to algorithms to the true ground state \cite{PhysRevC.108.L031306}.
In this Hamiltonian the total $J, J_{\rm{z}}$ and 
total translation/mirror symmetry each reduce the space by a factor proportional to
the size of the lattice. These impacts will remain very important for larger lattices.

\begin{figure*}[h]
    \centering
    \includegraphics[width=0.55\linewidth]{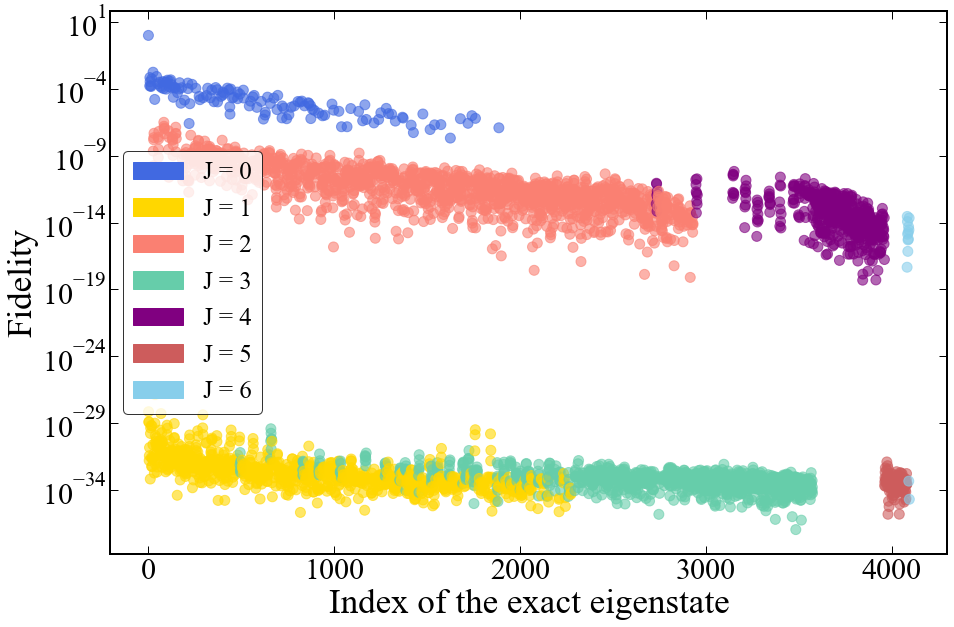}
    \caption{Fidelity of the prepared state obtained using the VQE algorithm with all-to-all connectivity, with respect to the reference ground state of the Neutrino model. The fidelity is shown on a logarithmis scale, where each total spin subspace $S$ is indicated by a different color.}
    \label{fig:neutrino}
\end{figure*}

In Fig. \ref{fig:neutrino} we show the probability of all eigenstates in the variational state for the neutrino case. The overlap is by far the largest with the true ground state, the other $J=0$ states
have probabilities of $10^{-5}$ to $10^{-4}$.
The spin $2$ states have the next highest probabilities after $11$ projection steps, around $10^{-9}$. The $J=2,4$ states have similar amplitudes, but there are fewer of those in this finite system because we are also projecting on $J_{\rm{z}}$. The $J=1$ states and all $J=3$ or higher odd $J$ states  have probabilities around $10^{-30}$ after even $11$ steps.  The trial state has very little $J=$ odd contamination, the states with $J=2,4,6$ are higher. These
probabilities 
could be reduced dramatically with a few more iteration steps if desired.

\section{Conclusions}

We have studied two quantum spin Hamiltonians
with variational quantum algorithms.  These two problems have different spectral gaps, spectral densities and symmetries.  In both cases projecting to the correct symmetries and then preserving them, at least approximately, with the variational ansatz is extremely effective in improving the efficiency of the variational method.  Additional studies of different spin problems or continuum Fermin problems represented in a set of basis states would be very valuable.

The variational solutions obtained remain approximate and could be further refined using time-projection techniques, such as QPE and related algorithms. By constraining the variational states in this algorithm to the correct symmetries, the relevant spectral gaps are increased—sometimes substantially. Since projection techniques require time evolution proportional to the inverse of the gap, combining variational algorithms with strong symmetry enforcement and projection methods can be especially powerful. Moreover, because variational steps can be interleaved with time evolution, even more efficient algorithms may be achievable for a wide range of important quantum state preparation tasks. Further investigations into strategies for reducing circuit depth—and thereby lowering noise sensitivity—would be highly beneficial.

In the neutrino model, projection into the $J=0$ subspace improved the energy from $-24.93$ to $-44.15$. This brought the state $\approx 80\%$ of the way towards the exact ground-state energy before any variational optimization. The corresponding fidelity with the exact ground state increased from $2.4\%$ to $56.7\%$. In the Heisenberg model, the spin projection step led to a smaller improvement from $-32.00$ to $-45.33$ in energy, and from $8.5\%$ to $29.6\%$ in fidelity.

The optimization of VQE further improved both systems. The neutrino model reaching $98.8\%$ fidelity while the Heisenberg model reaching $89.7\%$. By applying the additional projection to preserve translational and mirror symmetries increased the fidelity above $98\%$. These results confirm that symmetry preservation improves convergence and also produces states that are much closer to the true ground state.

For the $4 \times 3$ Heisenberg model, the full $2^{12} = 4096$-dimensional space is reduced to $924$ states by projecting onto $J_{\rm{z}}=0$, then to $132$ by restricting to $J=0$, and finally to only $9$ states when translational and mirror symmetries are also preserved. This reduction in dimensionality is accompanied by the increase in the effective spectral gap, from $\approx 7$ ($-58.95$ and $-51.81$) to nearly $28$ ($-58.95$ and $-46.72$), making projection algorithms more efficient.

These observations suggest that the symmetry projection is applicable to other many-body systems where the symmetry group can be efficiently implemented in a quantum circuit. The combination of symmetry-preserving variational ansatz and projection steps can yield highly accurate results with shallower circuits, which makes the approach well-suited for near-term quantum hardware. Extensions of this work to larger lattices, other spin models, and fermionic problems would help to further assess the scalability and generality of this strategy.







\section*{Acknowledgements}
I.M. acknowledges financial support from the Czech Academy of Sciences
(the {\it Praemium Academiae} awarded to Martin Friák, from
IPM in Brno, Czech Republic).  She was also partially supported by the U.S. DOE through a quantum computing program sponsored by the LANL Information Science \& Technology Institute. This work was carried out under the auspices of the National Nuclear Security Administration of the U.S. Department of Energy at Los Alamos National Laboratory under Contract No. 89233218CNA000001. IS and JC gratefully acknowledge support by the Advanced Simulation and Computing program.
 JC and DN also acknowledge the Quantum Science Center
 for partial support of their work on this project. 
Computational resources were provided by
the Ministry of Education, Youth and Sports of the Czech Republic under
the Projects e-INFRA CZ (ID:90254) at the IT4Innovations National
Supercomputing Center (computational time project assigned to Martin Friák, from IPM in Brno, Czech Republic). 

\bibliography{bibliography}

\end{document}